\pdfoutput=1
\documentclass[9pt, conference]{IEEEtran}
\IEEEoverridecommandlockouts
\usepackage{mathtools,amsmath,amsthm,amssymb,amsfonts}
\usepackage{graphicx}
\usepackage{textcomp}
\usepackage{xcolor}
\usepackage{csquotes}
\usepackage{booktabs}
\usepackage{caption,subcaption}
\usepackage[backend=biber,style=ieee,sortlocale=en-US,url=true,doi=false,date=year,minnames=3,maxnames=6,isbn=false, sortcites, dashed=false]{biblatex}
\usepackage{siunitx}
\usepackage{multirow}

\newtheorem{example}{Example}

\newtheorem{definition}{Definition}

\newcommand*{\atleast}{\textup{\textsc{AtLeast}}}
\newcommand*{\atmost}{\textup{\textsc{AtMost}}}

\usepackage[colorlinks=true, allcolors=black]{hyperref} 

\hypersetup{
	pdftitle={Using Boolean Satisfiability for Exact Shuttling in Trapped-Ion Quantum Computers},
	pdfsubject={29th Asia and South Pacific Design Automation Conference (ASP-DAC) 2024},
    pdfauthor={Daniel Schoenberger, Stefan Hillmich, Matthias Brandl, Robert Wille}
}

\begin{document}

\title{Using Boolean Satisfiability for Exact Shuttling\\in Trapped-Ion Quantum Computers\vspace{-0.3em}}
\author{\begin{tabular}[t]{c@{\extracolsep{4em}}c@{\extracolsep{4em}}c@{\extracolsep{4em}}c}
    \large Daniel Schoenberger\textsuperscript{1} & Stefan Hillmich\textsuperscript{2} & Matthias Brandl\textsuperscript{3} & Robert Wille\textsuperscript{1,2}
    \end{tabular}\\\\

\small
\textsuperscript{1} Chair for Design Automation, Technical University of Munich, Germany\\
\textsuperscript{2} Software Competence Center Hagenberg GmbH, Austria\\
\textsuperscript{3} Infineon Technologies AG, Germany\\
\href{mailto:daniel.schoenberger@tum.de}{daniel.schoenberger@tum.de}, \href{mailto:stefan.hillmich@scch.at}{stefan.hillmich@scch.at}, \href{mailto:matthias.brandl@infineon.com}{matthias.brandl@infineon.com}, \href{mailto:robert.wille@tum.de}{robert.wille@tum.de}\\
\url{https://www.cda.cit.tum.de/research/quantum/}
}

\maketitle

\begin{abstract}
    Trapped ions are a promising technology for building scalable quantum computers. Not only can they provide a high qubit quality, but they also enable modular architectures, referred to as \emph{Quantum Charge Coupled Device}~(QCCD) architecture. Within these devices, ions can be shuttled (moved) throughout the trap and through different dedicated zones, e.g., a memory zone for storage and a processing zone for the actual computation. However, this movement incurs a cost in terms of required time steps, which increases the probability of decoherence, and, thus, should be minimized. In this paper, we propose a formalization of the possible movements in ion traps via Boolean satisfiability. This formalization allows for determining the \emph{minimal} number of time steps needed for a given quantum algorithm and device architecture, hence reducing the decoherence probability. An empirical evaluation confirms that---using the proposed approach---minimal results (i.e., the lower bound) can be determined for the first time. An open-source
implementation of the proposed approach is publicly available at \url{https://github.com/cda-tum/mqt-ion-shuttler}.
\end{abstract}

\vspace{-0.3em}

\section{Introduction}\label{sec:introduction}
\vspace{-0.3em}
Quantum computing as a new paradigm promises to solve certain problems which are computationally intractable on classical computers. 
Famous examples include Shor's algorithm to factorize integers~\cite{DBLP:conf/focs/Shor94} and Grover's search for unstructured data~\cite{DBLP:conf/stoc/Grover96}, but also problems in quantum chemistry can profit from simulation in an actual quantum computer~\cite{PhysRevX.8.011044}.
The advantage of quantum computing lies in the exploitation of quantum mechanical effects~\cite{DBLP:books/daglib/0046438}, such as \emph{superposition}, where a quantum state can assume a combination of basis states, and \emph{entanglement}, where qubits can lose their locality and can no longer be described individually.
This potential also led companies to heavily invest in research towards quantum computing, as witnessed by IBM, Alphabet (Google), Microsoft, Rigetti, AQT, Infineon Technologies, and IonQ.

Among the possible physical technologies, such as superconducting quantum computers~\cite{Clarke2008}, neutral atom quantum computers~\cite{Henriet2020quantumcomputing, schmid2023computational}, and optical quantum computers~\cite{Knill2001}, trapped-ion quantum computers are one of the most promising candidates to show quantum advantage in the foreseeable future~\cite{Bruzewicz_2019}.
This is due to their ability to scale from smaller modules in an approach termed \emph{Quantum Charge Coupled Device} (QCCD,~\cite{Kielpinski2002,Pino2021}).
In this architecture, the qubits are encoded into ions that are trapped by electromagnetic fields. By manipulating these fields, the ions can be moved on the architecture.

However, the scaling of trapped-ion quantum computers requires corresponding tooling support to exploit the full potential.
Without proper support, there is the possibility that powerful trapped-ion quantum computers will be available but there will be no means to use that power.
Indeed, this holds true for all quantum computing technologies.
For ion traps in particular, efficiently moving (i.e.,~\mbox{\emph{shuttling}}) the ions on a QCCD architecture is essential because of the potential loss of quantum information over time due to decoherence.
This makes determining exact schedules for the movement paramount for efficient computations in trapped-ion quantum computers.

First solutions addressing this problem have been proposed, e.g.,~in~\cite{durandau2022automated, 9138945}. 
However, the considered architectures do not cover a large part of possible QCCD architectures and all solutions rely on heuristics, hence, cannot guarantee exactness. 
Moreover, generating \emph{exact} (i.e., minimal) solutions requires traversing a huge search space and, hence, constitutes a computationally expensive design task (a reason why previously proposed works resorted to heuristics in the first place). Boolean satisfiability, i.e., formulating the problem in a symbolic fashion as a \emph{satisfiability} (SAT) problem and using dedicated solvers afterward, promises to be an efficient approach. In fact, through sophisticated reasoning strategies, modern SAT solvers manage to cope with huge search spaces as demonstrated in numerous (classical) design automation tasks (see e.g., ~\cite{DBLP:reference/mc/BiereK18,DBLP:conf/iccad/Brand93,DBLP:conf/iccad/EggersglussWD13,DBLP:conf/date/GebregiorgisT19,DBLP:conf/fmcad/KaufmannBK19,DBLP:journals/tcad/Larrabee92,DBLP:conf/fdl/WilleGHD09, DBLP:journals/tcad/HaaswijkSMM20,DBLP:journals/fuin/KhomenkoKY06}) and, recently, also got utilized for quantum computing~\cite{Wille_2019, DBLP:conf/sat/BerentBW22, DBLP:journals/corr/abs-2301-11935, peham2023depthoptimal}.
In this paper, we are investigating whether this promising approach can also be utilized for determining an exact shuttling in trapped ion quantum computers.

To this end, we propose a symbolic encoding of the shuttling problem that can be used to employ SAT solvers in order to determine exact solutions.
More precisely, we first define an abstraction of general QCCD~architectures as a graph that captures information about the system state that is necessary for solving the shuttling problem---most importantly the position of ion chains at given times.
Based on this abstraction, we subsequently propose a symbolic encoding of the system state with Boolean variables.
This encoding is then constrained 
to ensure (i)~that only valid system states can be generated and (ii)~that transitions between system states only include changes that are possible on the hardware.
By employing a SAT~solver afterward, 
we finally attain the desired exact solution.

Empirical evaluations confirm the efficacy of the proposed approach. For the first time, exact (i.e., minimal) solutions for the shuttling problem for trapped-ion quantum computers are determined. Although this is only possible for small architectures (after all, the problem remains computationally expensive), this provides the first lower bounds for this problem. Moreover, the obtained exact results allow for a more sophisticated analysis and design exploration of architectures for this promising technology.
An open-source implementation of the proposed approach is publicly available at \url{https://github.com/cda-tum/mqt-ion-shuttler}.

The remainder of this paper is structured as follows:
\autoref{sec:background} provides background on trapped-ion quantum computers and QCCD architectures.
\autoref{sec:exact-shuttling} motivates the problem and outlines the proposed solution.
\autoref{sec:symbolic-formalization} details the symbolic state descriptions and the constraints to ensure only valid states.
\autoref{sec:experiments} summarizes the obtained results. 
Finally, \autoref{sec:conclusion} concludes the paper.

\section{Background}\label{sec:background}
In this section, we briefly review the trapped-ion quantum computing technology and the challenges that have to be addressed to realize scalable devices. This way, we motivate the considered problem that is abstracted and solved in the following sections. For a more detailed description, the interested reader is referred to the provided references.

\subsection{Trapped-Ion Quantum Computing}
\label{sec:ion-trap-qc}

The main idea of trapped-ion quantum computers~\cite{PhysRevLett.74.4091,PhysRevLett.113.220501,Debnath2016,9138945} is to use ions as entities for qubits whose states are manipulated by electromagnetic interactions in the optical or the microwave domain. To this end, ions are confined and suspended in free space using electromagnetic fields.
A trap can hold a chain of multiple ions confined in an electric potential. 
The potential is created by a combination of radio-frequency and quasi-static electric fields produced by control electronics. The individual ion chains have been coined \emph{ion registers}, because they may be used similarly to registers in classical computers.

\begin{figure}
    \centering
    \begin{subfigure}{0.45\linewidth}
        \centering
        \includegraphics[trim=0 2mm 0 0, clip, width=.9\linewidth]{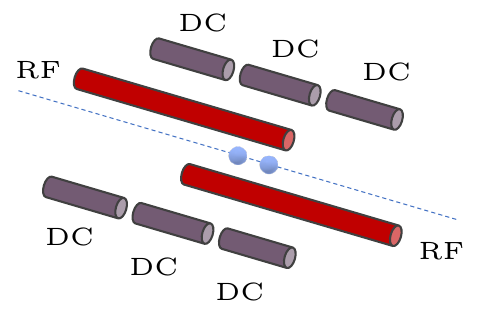}
        \caption{3D Paul trap}
        \label{fig:single-linear-trap-3d}
    \end{subfigure}\qquad%
    \begin{subfigure}{0.45\linewidth}
        \centering
        \includegraphics[trim=0 3mm 0 0, clip, width=.6\linewidth]{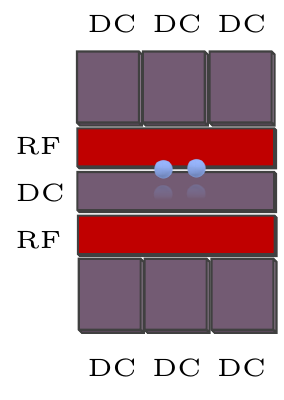}
        \caption{2D surface trap}
        \label{fig:single-linear-trap-2d}
    \end{subfigure}
    \caption{An illustration of two possible linear trap realizations. The combination of radio-frequency (RF) and quasi-static (DC) electric fields produced by control electronics creates a potential that confines the ions (blue).}
    \label{fig:single-linear-trap}
\end{figure}

\begin{example}
    \autoref{fig:single-linear-trap} shows two realizations of a trap that holds a single chain. 
    We refer to this as a \emph{single linear trap}.
    In both realizations, the ions are held in an electric field generated by radio-frequency (RF, red) and quasi-static (DC, purple) elements.
    In \autoref{fig:single-linear-trap-3d}, the trap is constructed as a \emph{3D linear Paul trap} with 3D control electronics.
    The same combination of control electronics can also be combined into a two-dimensional \emph{surface trap}, as depicted in \autoref{fig:single-linear-trap-2d}.

\end{example}

While trapping all ions in one trap is feasible for small quantum computers, physical limitations prevent this approach from scaling to larger qubit numbers that are needed for practical quantum algorithms~\cite{Pino2021}.
Increasing the number~$N$ of ions  in a trap leads to slower gate speed~$R_{\textrm{gate}}$ which approximates to $R_{\textrm{gate}} \sim \frac{1}{\sqrt{N}}$. Longer gate times give rise to different types of background errors---limiting the size of a practical single trap. As of now, trapped-ion quantum computers have been realized using up to tens of qubits~\cite{Brown2021}.

\subsection{Quantum Charge Coupled Device Architecture}
\label{sec:qccd}

These limitations can be addressed by building modular systems, which can hold multiple ion chains. 
A promising modular approach is called the \emph{Quantum Charge Coupled Device}~(QCCD) architecture~\cite{9138945}.  
The QCCD approach proposes to exploit the fact that ion chains can be split, merged, and moved through the system by applying different operating voltages to the control electrodes. The key idea of the QCCD approach is then to assign certain trap regions certain tasks. For example, ion chains that store quantum information may be placed into a zone dedicated to \emph{memory}, where ions are shielded from potential sources of decoherence. All quantum operations are then performed in a separated \emph{processing} zone that is specifically tailored toward efficient qubit interactions. A complete QCCD device may also include optimized regions like \emph{measurement} zones for the readout of qubit information and \emph{loading} zones for initializing new ions. In its simplest form, a QCCD device is constructed in a straight line to form a linear system. At every site, the system can trap one chain. A linear QCCD trap has been successfully realized in~\cite{Debnath2016,Pino2021}.
\begin{figure}[t]
    \centering
    \includegraphics[width=.6\linewidth]{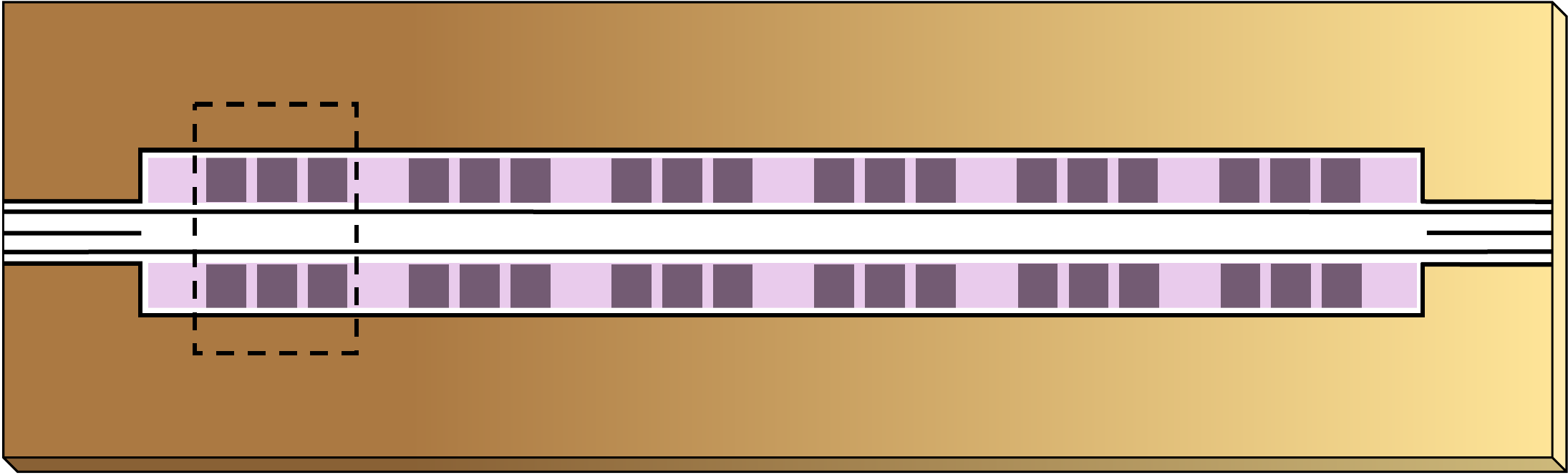}
    \caption{An illustration of a linear QCCD device. In this example, each site (indicated by a dotted line) consists of six control electrodes (purple). At every site exactly one ion chain can be trapped between the electrodes similar to the surface trap in \autoref{fig:single-linear-trap-2d}.}
    \label{fig:linear-demo}
\end{figure}

\begin{example}\label{ex:linear-trap}
    \autoref{fig:linear-demo} shows a schematic linear QCCD device. 
    Three control electrodes (purple) combined on each side of the horizontal axis represent one site of the linear trap. The ion registers can be held in between the control electronics and shuttle to neighboring sites.
\end{example}

In a linear trap, ion chains cannot move out of each other's way, thus slow interactions like chain reordering and reconfiguration are needed to address all ions. To tackle this problem, linear regions can be connected via junctions to form two-dimensional (2D) architectures. If an ion chain blocks the way of its neighbor chain, it can use a junction to simply move out of the way. 2D QCCD architectures can be constructed in a variety of ways. One may use only small linear regions and a large number of junctions, or instead, reduce the number of junctions and employ larger linear regions. 

\begin{figure}[t]
    \centering
    \begin{subfigure}{0.45\linewidth}
        \centering
        \includegraphics[width=\linewidth]{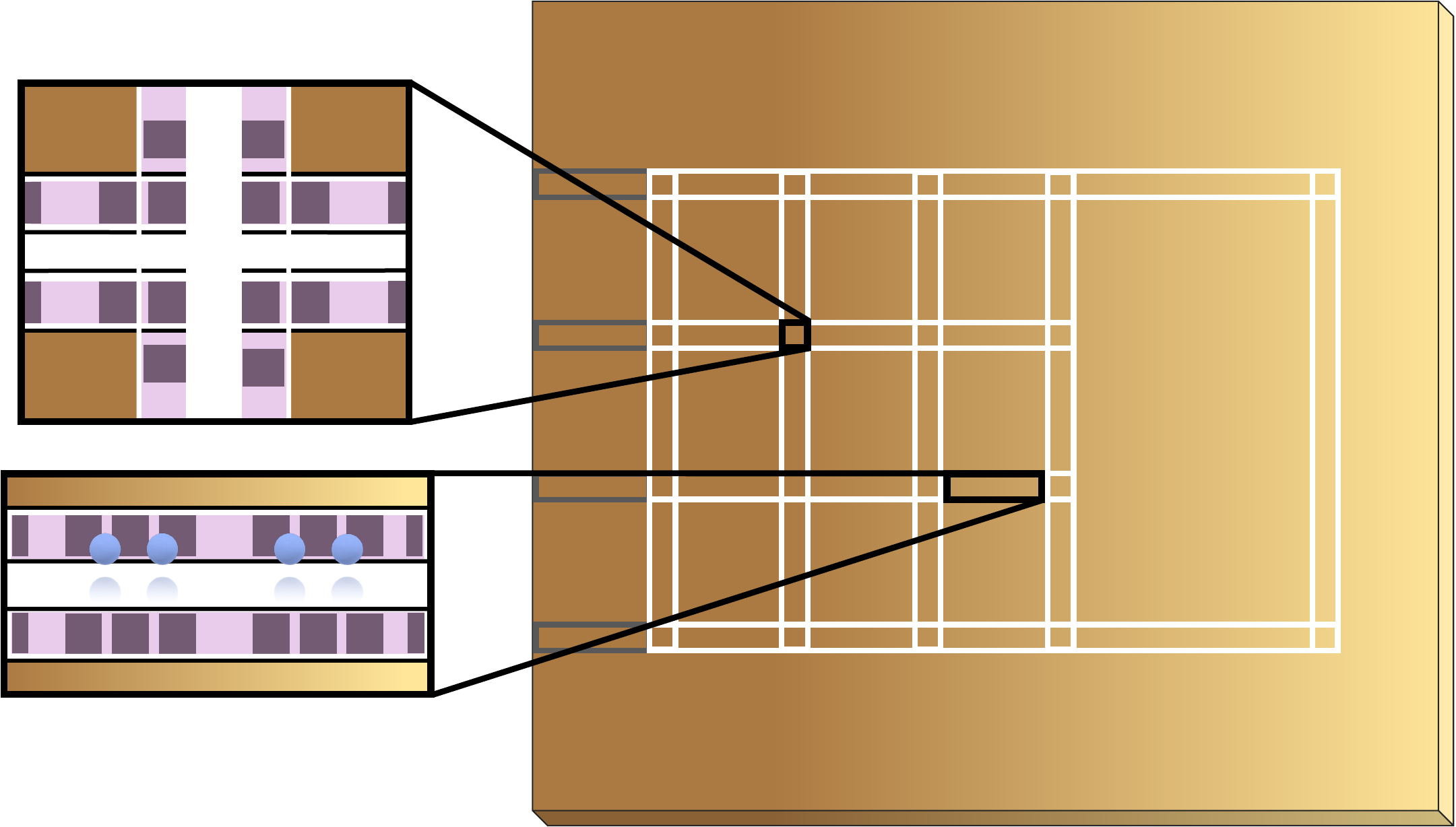}
        \caption{QCCD architecture}
        \label{fig:qccd-example}
    \end{subfigure}\qquad%
    \begin{subfigure}{0.45\linewidth}
        \centering
        \includegraphics[trim=0 23mm 0 0, clip, width=.8\linewidth]{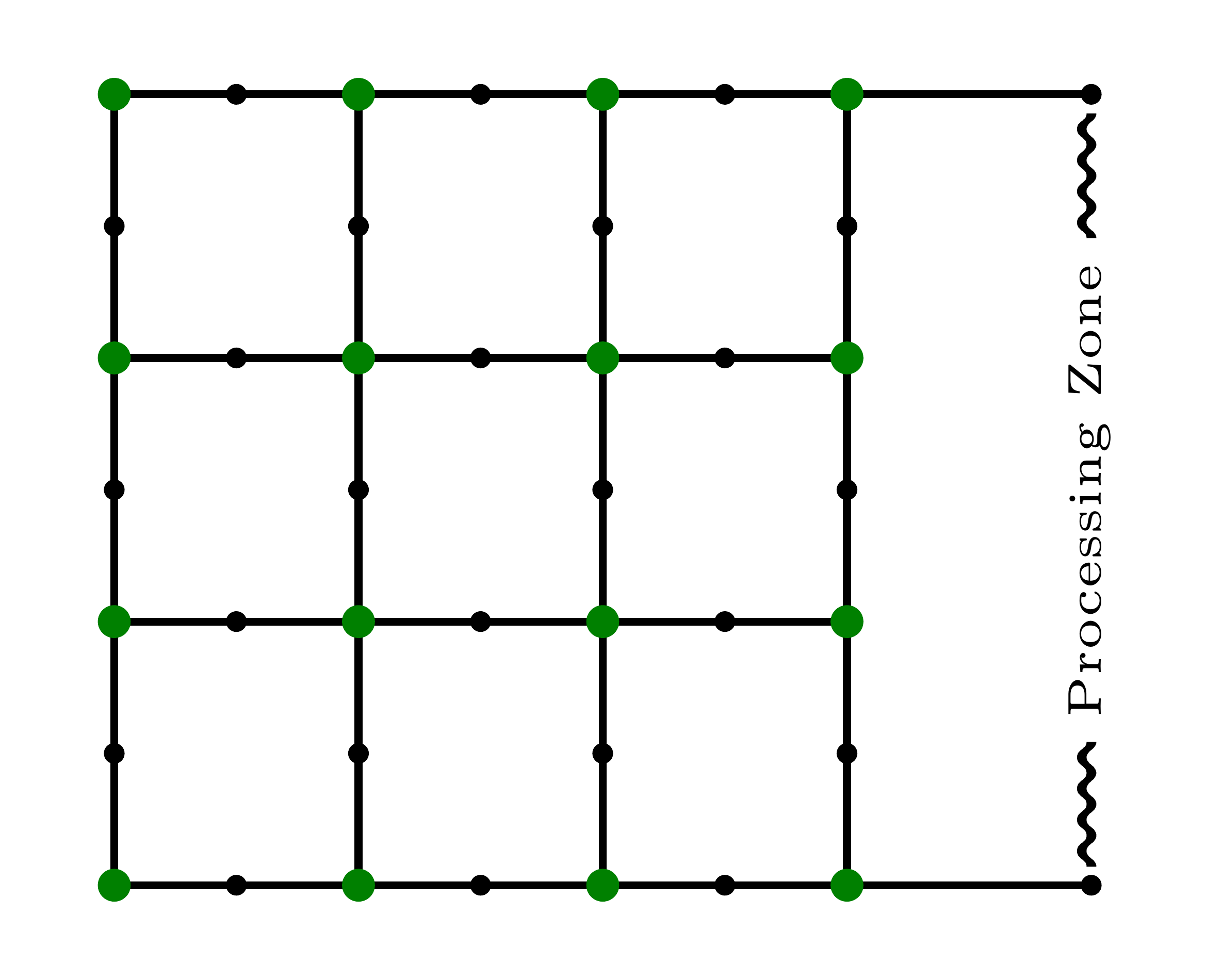}
        \caption{Corresponding Graph}
        \label{fig:graph-example}
    \end{subfigure}
    \caption{An illustration of a 2D QCCD architecture and its corresponding graph}
    \label{fig:qccd-to-graph}
\end{figure}

\begin{example}
    \autoref{fig:qccd-example} shows one possible layout of a 2D QCCD device. 
    Inner linear regions are connected via \enquote{X}-junctions, which produce a system in a grid structure. Each linear region can hold up to two chains. 
    In this architecture, the square grid on the left-hand side is dedicated as a memory zone which is connected to a linear processing zone on the right-hand side.
\end{example}

To execute a quantum algorithm on a QCCD device, the ions have to be shuttled to the correct zone at the right time. A quantum circuit determines in which order ions have to be moved from the memory zone to the processing zone. After they have been processed in the processing zone, the ions move back to the memory zone to store their acquired quantum information.

\subsection{Shuttling Costs}

Conducting the shuttling operations of ion chains in a QCCD architecture comes with certain costs.
Depending on the architecture, different amounts of energy have to be invested into the system to trap and move the ions. The more the ions move through the system, the more they collect energy through the acquisition of phonons~\cite{arxiv.1702.02583}. To avoid losing quantum information, ions must be cooled to or close to the ground state---often by a combination of Doppler and side-band cooling (see \cite{Bruzewicz_2019} for more information).
Integrating and scaling the required optical control elements still proves to be difficult. It is therefore crucial to find exact shuttling schemes that minimize the time ions have to be moved through the system.

\section{Exact Shuttling in QCCD Architectures}
\label{sec:exact-shuttling}

As described above, the execution time of a given algorithm is dictated by the time that the respective ion chains need to move from the memory zone to the processing zone and the required movement inside the memory zone to give way for other ion chains. 
Since the possible movement inside the processing zone is immediately given by the architecture, we put the focus of this paper on the movement inside the memory zone and at the interface to the processing zone.
In this section, we formulate these considerations as a discrete problem set, by representing the architecture of a memory zone and the interface to the processing zone in a QCCD device as an undirected graph. 
Time can then be discretized into time steps, where at every time step, one can describe the location of the ion chains on the graph by means of Boolean variables in a symbolic fashion. 
Based on this representation, we propose an encoding as a Boolean satisfiability problem to determine an exact, i.e., minimal, solution.

\subsection{Considered Problem}

The problem we are considering in this paper is determining an exact way of shuttling in a QCCD architecture given a quantum circuit.
As discussed before, the ion chains can not be split up or directly swap places within a memory zone, therefore, each chain is abstracted as one individual entity moving on a graph.
Further, since movement outside the memory zone only allows very limited choices, we focus our investigations on the memory zone. In the following, the abstraction of a QCCD architecture is referred to as a \emph{layout}.

\begin{definition}\label{def:notation}
    Consider a graph $G = (V, E)$ that represents the architecture of the memory zone and the interface to the processing zone on the QCCD device. %
    
    The set of nodes $V$ contains two different types: \emph{major nodes} representing junctions and \emph{minor nodes} separating adjacent traps.
    
    The set $E = \{e_0, \dots, e_k\}$ denotes the edges of the graph, representing all possible positions of the ion chains.
    There are two special edges: The (i)~\emph{outbound edge} for ion chains exiting the memory zone for processing in the processing zone and the (ii)~\emph{inbound edge} for ion chains returning to the memory zone. 
    Edges can connect two major nodes (when there is only one site between junctions), a major and a minor node, as well as two minor nodes.
    
    In addition to the fixed architecture, there is also the set of ion chains $C = \{i_0, \cdots, i_{l}\}$ which contains all ion chains considered in the given shuttling problem.
    One ion chain can contain multiple individual ions, where each ion carries the information for one qubit.
\end{definition}

\begin{example}
    Consider the QCCD architecture shown in \autoref{fig:qccd-example}.
    All regions between two junctions in the memory zone can hold up to two individual ion chains. 
    The corresponding graph is illustrated in \autoref{fig:graph-example} where the major nodes (green) form a grid of size $4\times 4$.
    Minor nodes (black) separate the two individual sites between two junctions. The edge of the processing zone is labeled and connected via one inbound and one outbound edge to the memory zone.
\end{example}

On this graph-based abstraction, we can describe the state of the memory zone at any time.
Discretizing the time into individual \emph{time steps} enables the inclusion of movement in the description without having to include hardware-dependent information. 
Each time step represents the system at a given time and, between time steps, ion chains can move through junctions.
More precisely, per transition between time steps, at most one ion chain may pass through a given junction as long as there is an available site it can move into \enquote{after} the junction.

While the underlying architecture constrains the possible ways an ion chain may move, the considered quantum algorithm dictates which ion chains have to be at what position at what time.
More precisely, since the considered quantum algorithm defines the sequence of gates and, hence, the sequence of qubits needed in the processing zone, the corresponding ion chains including those qubits have to move to the outbound edge and, after processing, return to the inbound edge and back into the memory zone.
This can be represented by a \emph{sequence} $S$ of ion chains derived from the respectively considered quantum algorithm. 

Based on all that, the overall problem is defined as follows:
\begin{itemize}
    \item \textbf{Given:} A graph layout $G$ of a memory zone with a one-way connection to a processing zone and a quantum circuit as a sequence $S$.
    \item \textbf{Goal:} Determine the minimum amount of time steps $\hat{T}$ for shuttling to execute the circuit.
\end{itemize}

\subsection{General Idea}
\label{sec:general-idea}

The problem defined above, namely determining an exact shuttling schedule for trapped-ion quantum computers,
is an optimization problem that yields significant computational complexity. 
At the same time, it constitutes a classical combinatorial optimization problem.
In the past, SAT solvers have been proven very beneficial and efficient in tackling this complexity. Accordingly, in this work, we propose to 
exploit this power and, by this, provide an example of how quantum computing can benefit from the potential of SAT solving.
To this end, we briefly review the basics of Boolean satisfiability to keep the work self-contained.

\begin{definition}
    Let $\Phi$ be a Boolean function. Then, the \emph{Boolean satisfiability} problem is to determine an assignment to the variables of $\Phi$ such that $\Phi$ evaluates to 1 or to prove that no such assignment exists.
\end{definition}

\begin{example}
    Let $\Phi = (x_1 \lor x_2 \lor x_3)\land(x_1 \lor x_3)\land(x_2 \lor x_3)$. Then, $x_1 = 1$, $x_2 = 1$, and~$x_3 = 1$ is a satisfying assignment for $\Phi$. %
\end{example}

SAT is one of the central NP-complete problems. 
Despite this complexity, today's SAT solvers incorporate powerful solving strategies~\cite{DBLP:series/faia/336}.
They can handle instances composed of hundreds of thousands of variables and, hence, found applications in automatic test pattern generation~\cite{DBLP:journals/tcad/Larrabee92,DBLP:conf/iccad/EggersglussWD13,DBLP:conf/date/GebregiorgisT19}, logic synthesis~\cite{DBLP:journals/tcad/HaaswijkSMM20,DBLP:journals/fuin/KhomenkoKY06}, verification~\cite{DBLP:conf/iccad/Brand93,DBLP:conf/fmcad/KaufmannBK19}, and more. While they have hardly been exploited in the domain of quantum computing yet, first promising SAT-based solutions recently have been proposed, e.g., in~\cite{DBLP:journals/corr/abs-2301-11935,DBLP:conf/sat/BerentBW22}. In this work, we are aiming to solve another important problem from this domain utilizing SAT---working towards establishing SAT-based solutions in quantum computing and, by this, repeating the success SAT had in classical computing for this new domain.

In order to utilize SAT solvers for the considered problem, we first translate the given optimization problem into a sequence of decision problems.
More precisely, the question of \enquote{What is the minimum amount of time steps $\hat{T}$ for shuttling to execute the circuit?} is deconstructed into a sequence of \enquote{Can the circuit be executed with a shuttling procedure of $T$ time steps}.
An exact and, thus, minimal solution is then guaranteed by starting with~$T=1$ and increasing~$T$ by 1 whenever it could be shown that no valid solution with $T$ time steps exists.
With this approach, the first value of $T$ for which a satisfying solution is found is also the exact solution which requires $T=\hat{T}$ time steps.
This can then be formulated as a SAT instance~$\Phi$ in which all possible solutions are formulated in a symbolic fashion, i.e.,~through variables representing the problem and constraints enforcing the validity of the solution. Details on this encoding are provided in the next section.

\section{Encoding of the Shuttling Problem\\as an SAT Instance}
\label{sec:symbolic-formalization}

In this section, we propose an encoding of the previously introduced graph-based abstraction into a Boolean satisfiability problem.
To this end, we first provide a description of the variables which symbolically represent all possible system states; followed by constraints that ensure consistency and only valid transitions between two states.

\subsection{Symbolic State Description}

In order to symbolically encode all possible states in a QCCD architecture in the Boolean function $\Phi$, the following variables are introduced:
\begin{definition}
    Consider again \autoref{def:notation}.
    The Boolean variables employed to describe the state of the system are denoted $x^t_{e, i} \in\{0,1\}$ with $0 \leq t \leq T$, 
    $i \in C$, and $e \in E$.
    Each such variable represents whether there is an ion chain $i$ present on site $e$ at time step $t$.
    The value 0 (\enquote{false}) means absent, whereas 1 (\enquote{true}) means present.
\end{definition}

\begin{figure}[t]
	\centering
	\begin{subfigure}[b]{0.4\linewidth}
		\centering
		\includegraphics[trim=0 23mm 0 20mm, clip, width=\linewidth]{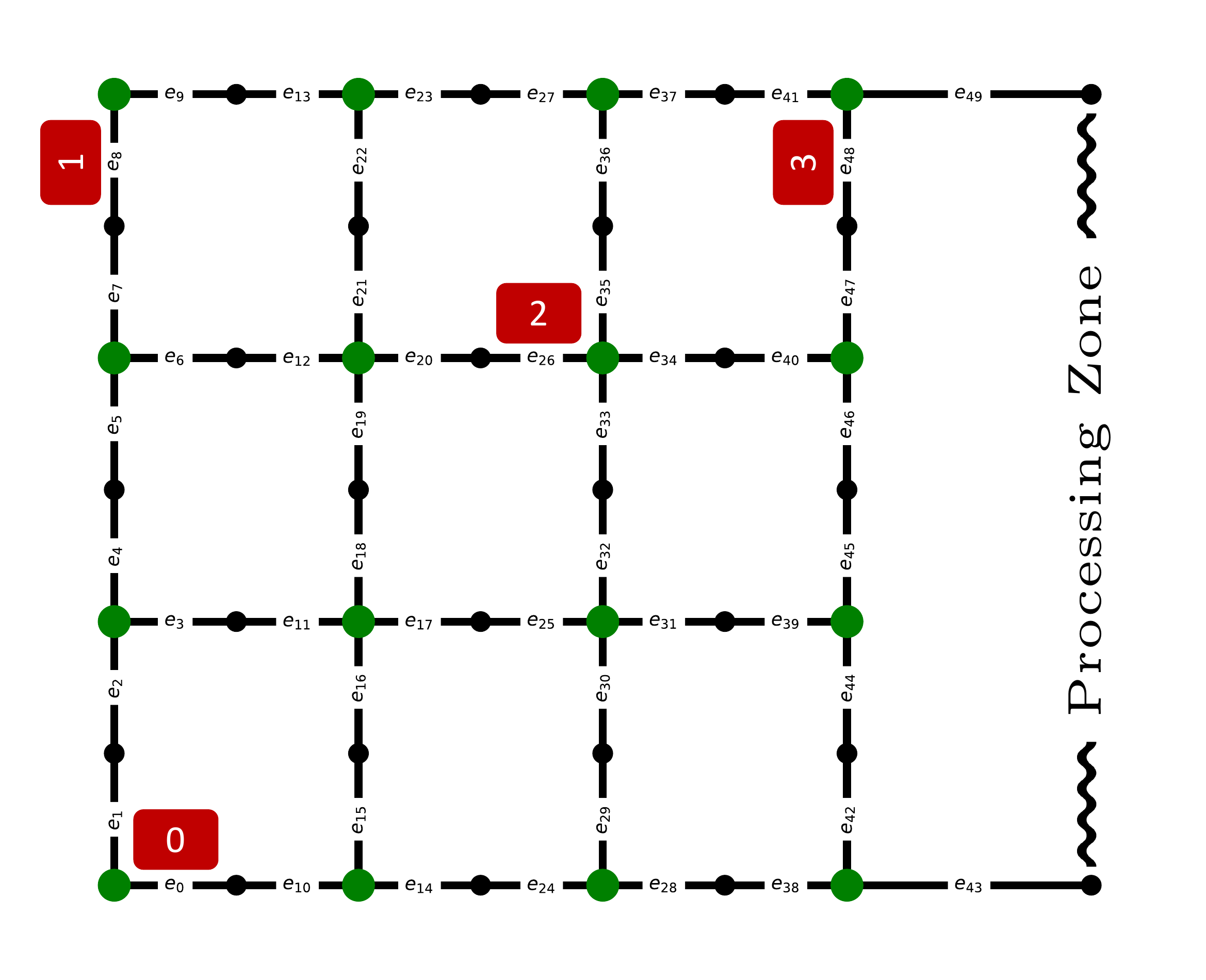}
		\caption{Time step 0}
		\label{fig:Variables-initial}
	\end{subfigure}\hfill%
	\begin{subfigure}[b]{0.55\linewidth}
		\centering
		\resizebox{\linewidth}{!}{\begin{tabular}{@{}llll@{}}
			ion chain 0 & ion chain 1 & ion chain 2 & ion chain 3 \\\\
			$\mathbf{x^0_{e_0, i_0}=1}$ & $x^0_{e_0, i_1}=0$ & $x^0_{e_0, i_2}=0$ & $x^0_{e_0, i_3}=0$ \\
			$x^0_{e_1, i_0}=0$ & $\vdots$ & $\vdots$ & $\vdots$ \\
			$\vdots$ & $\mathbf{x^0_{e_8, i_1}=1}$ & $\mathbf{x^0_{e_{26}, i_2}=1}$ & $\vdots$ \\
			$\vdots$ & $\vdots$ & $\vdots$ & $\mathbf{x^0_{e_{48}, i_3}=1}$ \\
			$x^0_{e_{49}, i_0}=0$ & $x^0_{e_{49}, i_1}=0$ & $x^0_{e_{49}, i_2}=0$ & $x^0_{e_{49}, i_3}=0$
		\end{tabular}}
		\caption{Encoding of the state at time step 0}
        \label{fig:Encoding-initial}
	\end{subfigure}
	\caption{Correspondence between the abstraction of a state and its encoding in Boolean variables.}
\end{figure}

\begin{example}
    Consider the $4 \times 4$-grid shown in \autoref{fig:Variables-initial} which represents the initial time~step of a possible shuttling problem. 
    The edges $e_0, e_8, e_{26}$, and $e_{51}$ are occupied by the ion chains $i_0, i_1, i_2$, and~$i_3$, respectively. 
    All other edges are empty, including the inbound and outbound edge interfacing the processing zone (which are always considered to be empty at the beginning).
    The resulting variables of this configuration for the time step~$t=0$ are shown in \autoref{fig:Encoding-initial}.
\end{example}

Having all these variables, the positions of all present ion chains can be described at any time step $t$. 
All time steps $0 \leq t \leq T$ combined, this eventually represents a shuttling sequence on the given grid. 
However, passing those variables without any further constraints to a SAT solver, obviously yields an arbitrary assignment and, hence, an arbitrary (and most likely invalid) shuttling sequence. To prevent that, additional constraints are enforced which are described next.

\subsection{Validity Constraints}

Given the symbolic encoding of the system state with the variables~$x^t_{e, i}$ from \autoref{def:notation}, we now provide constraints to ensure that assignments to the variables cannot result in an invalid system state.
For each constraint, we provide a textual description to provide an intuition as well as a formal description.
To build the function $\Phi$, the individual constraints are combined as a logical conjunction, i.e.,~the \textsc{And} operation. Furthermore, we define the following functions to increase the readability of constraints:
\begin{definition}\label{def:atleast-atmost}
    To improve the readability of the constraints, the functions \atleast{} and \atmost{} are used, which, given a set $M$ of Boolean variables, are defined as:
    {\footnotesize
    \begin{align*}
        \atleast(M) = \sum_{m \in M}(m) \geq 1 &&
        \atmost(M) = \sum_{m \in M}(m) \leq 1
    \end{align*}
    }
    Regarding the summation, \emph{true} values are treated as $1$ and \emph{false} values are treated as $0$.
    These functions (also termed \emph{cardinality constraints}) are provided by common SAT solvers and transparently translated into basic Boolean operations~\cite{10.1007/978-3-540-45193-8_8}.
\end{definition}

The first set of constraints ensures that the description at a single time step is valid:

Since we consider a trapped-ion quantum computer, each and every ion chain $i \in C$ in the model has to occupy exactly one site (modeled as edge $e \in E$) in each time step $t$, i.e.,
    {\footnotesize
    \begin{align*}
        \bigwedge_{0 \leq t \leq T} \bigwedge_{i \in C} \bigg(\atmost\Big(\big\{x^t_{e, i}\big\}_{e \in E}\Big) \wedge \atleast\Big(\big\{x^t_{e, i}\big\}_{e \in E}\Big)\bigg).
    \end{align*}
    }

In correspondence to the previous constraint, the following constraint ensures that each site can only hold zero or one ion chain in any given time step (in the case of the inbound edge $e_{in}$, this is relaxed to at most two ion chains, which allows us to simulate the presence of two ion chains in the processing zone), i.e.,
    {\footnotesize
    \begin{align*}
        \bigwedge_{0 \leq t \leq t_T} \bigwedge_{e \in E} \atmost\left(\left\{x^t_{e,i}\right\}_{i \in C}\right).
    \end{align*}
    }

After constraining the variables to only assume valid values regarding individual time~steps, the following section adds constraints to ensure the SAT solver only assigns valid values with respect to possible transitions (and, hence, movements) between time~steps.

\subsection{Movement Constraints}

\begin{figure}[t]
\centering
    \begin{subfigure}{0.43\linewidth}
      \centering
      \includegraphics[width=1.1\linewidth]{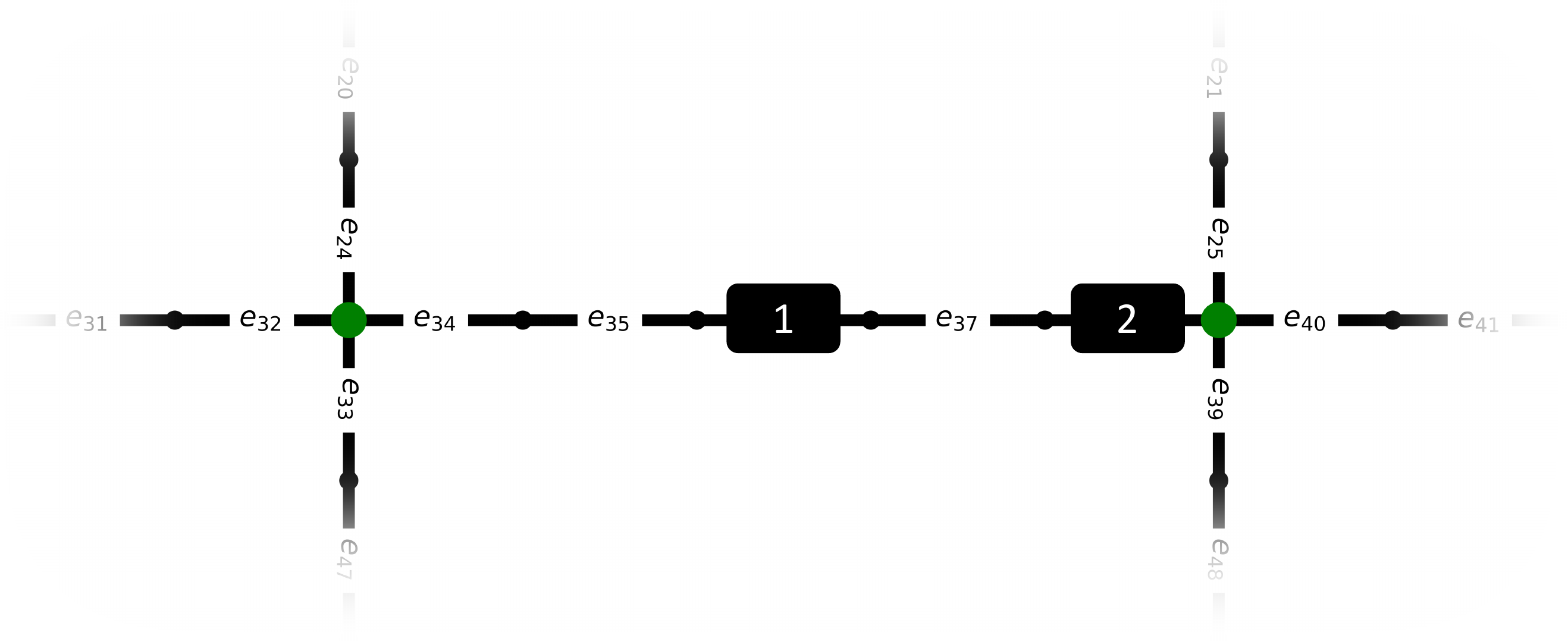}
      \caption{Time step $t$}
      \label{fig:Variables-movement}
    \end{subfigure}\hfill%
    \begin{subfigure}{0.54\linewidth}
    \centering
	\resizebox{\linewidth}{!}{\begin{tabular}{lll}
            time step $t$ & $x^t_{e_{36}, i_1}=1$ & $x^t_{e_{38}, i_2}=1$ \\
            time step $t+1$ & $x^{t+1}_{e_{36}, i_1}$ & $x^{t+1}_{e_{38}, i_2}$ \\
             & $x^{t+1}_{e_{37}, i_1}$ & $x^{t+1}_{e_{37}, i_2}$ \\
             & $x^{t+1}_{e_{35}, i_1}$ & $x^{t+1}_{e_{25}, i_2}$ \\
             & $x^{t+1}_{e_{24}, i_1}$ & $x^{t+1}_{e_{40}, i_2}$ \\
             & $x^{t+1}_{e_{32}, i_1}$ & $x^{t+1}_{e_{39}, i_2}$ \\
             & $x^{t+1}_{e_{33}, i_1}$ & \\
		\end{tabular}}
		\caption{Possible moves in symbolic encoding}
        \label{fig:moves-encoding}
    \end{subfigure}

    \begin{subfigure}{0.43\linewidth}
      \centering
      \includegraphics[width=1.1\linewidth]{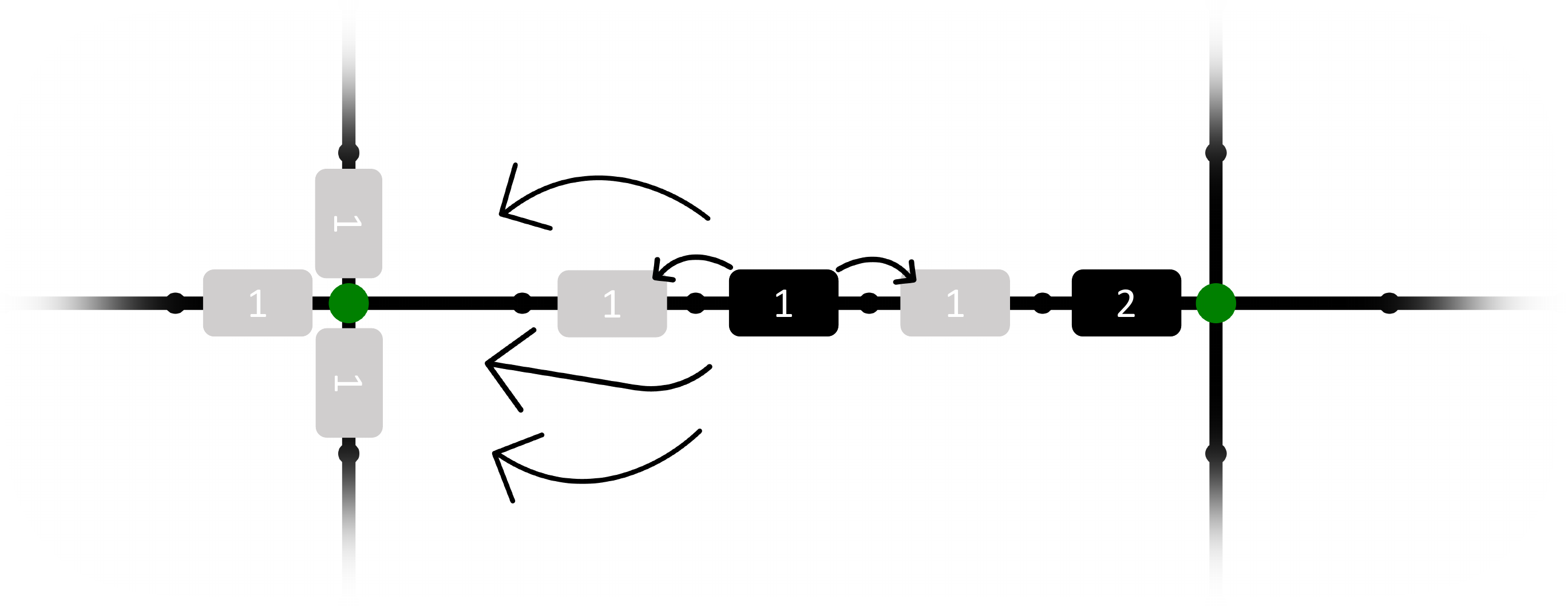}
      \caption{Possible moves of chain $i_1$}
      \label{fig:movement1}
    \end{subfigure}\hfill%
    \begin{subfigure}{0.47\linewidth}
      \centering
      \includegraphics[width=\linewidth]{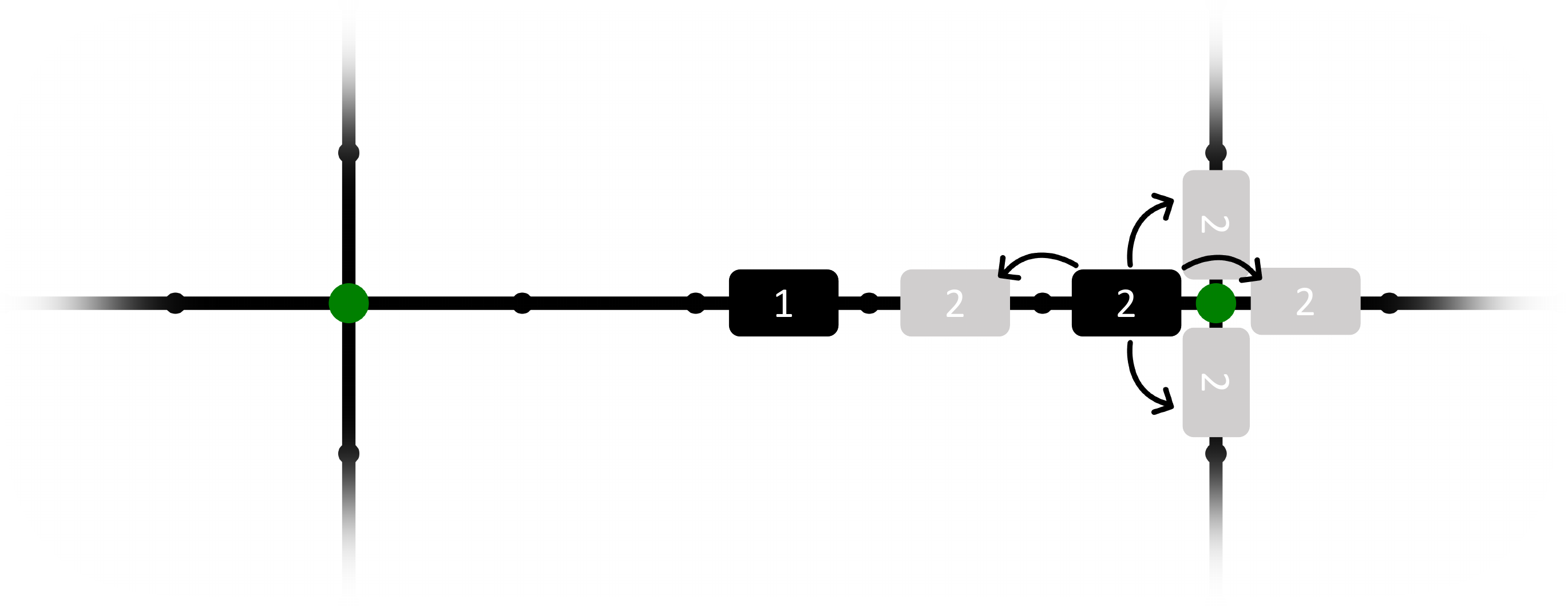}
      \caption{Possible moves of chain $i_2$}
      \label{fig:movement2}
    \end{subfigure}\hfill%
    \caption{Visualization and encoding of possible movement of two ion chains $i_1$ and $i_2$ in one time step. Current ion chain positions at time step $t$ are indicated as black boxes, possible positions at time step $t+1$ as grey boxes.}
    \label{fig:moves_visualization}
\end{figure}

Movement constraints ensure that only valid changes are assigned to the system in the transition from one time step to the next.

To formulate the movement of ion chains, we introduce the notations of \emph{neighbor edges}~$N(e)$ of an edge $e\in E$ and the \emph{path edges}~$P(e, e^*)$ between two edges $e$ and $e^*$. More precisely, $N(e)$ denotes edge $e$ and all edges that are directly connected to it. Additionally, $N^*(e)$ also includes all edges, that follow after the two nearest junctions of $e$. The set of path~edges~$P(e, e^*)$ contains all edges that are part of the shortest path in the graph between $e$ and $e^*$. \autoref{fig:moves_visualization} provides a visualization of $N^*(e)$. During the transition from one time step~$t$ to the next one $t+1$, each ion chain is allowed to stay at its current edge $e$ or move to one of the edges $e^*$ given by $N^*(e)$, if the path $P(e, e^*)$ in between $e$~and~$e^*$ is not occupied at time step $t$ by any other ion chain $i' \in C \setminus \{i\}$. Ion chains are always free to move to directly connected edges, e.g., edges of $N(e)$, since there are no edges between them. The corresponding encoding of possible moves for the ion chains in \autoref{fig:Variables-movement} is given in \autoref{fig:moves-encoding}. This approach allows every chain to move past one junction, given the path leading to this junction is not blocked, and also preserves the order of chains that are positioned in between two junctions.
    This leads to
    {\footnotesize
    \begin{align}
        \bigwedge_{0 \leq t \leq T} \bigwedge_{i \in C} \bigwedge_{e \in E } \hspace{-2pt} \Bigg(x^t_{e, i} \Rightarrow \hspace{-12pt} \bigvee_{e^* \in N^*(e)} \hspace{-3pt} \bigg( x^{t+1}_{e^*, i} \wedge \hspace{-10pt} \bigwedge_{e' \in P(e, e^*)} \bigwedge_{i' \in C \setminus \{i\}} \hspace{-10pt} \neg x^t_{e',i'} \bigg) \Bigg).
        \label{eq:mov-constr1}
    \end{align}
    }

As described before, it is not possible that ion chains are moved past each other in a memory zone. Furthermore, a junction can only shuttle one ion chain at a time. To enforce this in our encoding, we limit the number of ion chains that are allowed to pass over a node to~1. This includes every node in the layout, meaning all junctions and all minor nodes within two junctions. For this purpose, we introduce the set of edges $\textit{Sh}(v)$ which \emph{share} a node $v$, e.g., edges $e_1$ and $e_2$ share node $v$, if $v \in e_1$ and $v \in e_2$. Additionally, we denote the set of \emph{former} edges as $F(e)$. Former edges include all possible edges of an ion chain at $t-1$, given it occupies edge $e$ at $t$ and \autoref{eq:mov-constr1} is fulfilled.
    Overall, this leads to
    {\footnotesize
    \begin{align*}
        \bigwedge_{0 \leq t \leq T} \bigwedge_{v \in V} \atmost\Bigg(\bigg\{x^t_{e, i} \wedge \bigvee_{f \in F(e)} x^{t-1}_{f, i}\bigg\}_{i \in C; e \in \textit{Sh}(v)}\Bigg).
    \end{align*}
    }

We consider a one-way connection to the processing zone, i.e., ion chains that use the outbound edge $e_{out}$ have to move to the connected processing zone edge $e_{in}$ and back into the memory zone graph to a neighbor edge of the inbound edge $n \in N(e_{in})$. For simplicity, we split the corresponding encoding into three separate constraints:
    \begin{itemize}
        \item at the outbound edge, an ion chain has to move to the inbound edge at the next time step, i.e.,
            {\footnotesize
            \begin{align*}
                \bigwedge_{0 \leq t < T} \bigwedge_{i \in C} \big( x^t_{e_{out}, i} \Rightarrow x^{t+1}_{e_{in}, i} \big),
            \end{align*}
            }
        \item at the inbound edge, an ion chain was either in the inbound edge or in the outbound edge one time step before, i.e., 
            {\footnotesize
            \begin{align*}
                \bigwedge_{0 < t \leq T} \bigwedge_{i \in C} \big( x^t_{e_{in}, i} \Rightarrow x^{t-1}_{e_{out}, i} \vee x^{t-1}_{e_{in}, i} \big), \mbox{and}
            \end{align*}
            }
        \item a chain in the inbound edge has to either stay in $e_{in}$ or move to a neighbor edge $n \in N(e_{in})$ of $e_{in}$ (excluding $e_{out}$), i.e., 
            {\footnotesize
            \begin{align*}
                \bigwedge_{0 \leq t < T} \bigwedge_{i \in C} \bigg( x^t_{e_{in}, i} \Rightarrow \bigvee_{n \in N(e_{in}) \setminus \{e_{out}\}} x^{t+1}_{e_{in}, i} \vee x^{t+1}_{n, i} \bigg).
            \end{align*}
            }
    \end{itemize}

\begin{example}
    Consider the scenario in \autoref{fig:moves_visualization}. 
    \autoref{fig:Variables-movement} illustrates the system state at time step $t$ and \autoref{fig:moves-encoding} gives the possible positions of the considered ion chains in time step $t+1$ in terms of variables in the encoding.
    All possible moves of ion chain $i_1$ (ion~chain~$i_2$) are illustrated in \autoref{fig:movement1} (\autoref{fig:movement2}). Both ion chains can either stay at their current edge, move to their neighbor edge or move over a connected junction. Since they block the path of each other, only the moves passing the opposite junction are possible for each ion chain.
\end{example}

\subsection{Enforcing the Target Sequence}

The combination of the discussed validity and movement constraints allows a solving engine to determine valid configurations of ion chain positions that simulate the movement of ion chains in a QCCD device for a given amount of time steps. 
Since we want to determine  if a given sequence of ions can be shuttled in order to the processing zone in $T$ time steps, we introduce further constraints to enforce this goal. 
As discussed above, a quantum algorithm results in a sequence of qubits that have to be processed in the processing zone. 
Each qubit is mapped to an individual ion. 
Each ion is part of one ion chain. For clarity, we only consider ion chains that contain exactly one ion and therefore one qubit, i.e., the sequence of qubits is identical to the sequence of ion chains.
The generalization to ion chains with multiple ions is straightforward, since this only changes the resulting sequence. This can be handled with data preprocessing.
\begin{example}\label{ex:sequence_ion_chains}
    The sequence of qubits that have to be processed in the processing zone for a quantum Fourier transform with four qubits reads $[q_0; (q_0, q_1); (q_0, q_2); q_1; (q_1, q_2); q_3]$. For a given architecture with four ion chains, each holding exactly one ion, the resulting sequence of ion chains is given by: $[i_0; (i_0, i_1); (i_0, i_2); i_1; (i_1, i_2); i_3]$.
\end{example}

In order to formulate a constraint that enforces this sequence, we introduce helper variables that describe the elements of the sequence.
\begin{definition}
    We represent every step of the sequence as a Boolean variable $s^t_j \in \{0,1\}$  with $0 \leq t \leq T$ and $j \in C$. Each such variable represents one element of the sequence with index $j$ at time step $t$.
    The value 0 (\enquote{false}) indicates that the corresponding element is not executed, while 1 (\enquote{true}) means it is executed at this time step.
\end{definition}

Then, to make use of these helper variables, we need to connect them with the corresponding variables $x^t_{e, i}$. For example, if the variable of sequence element $j$ corresponding to ion chain $i_k$ is 1, variable $x^t_{e_{in}, i_k}$ must also be 1. This means, that if $s^t_j$ is true, the respective ion chains have to be in the inbound edge $e_{in}$ at $t$, which confirms that they have been moved to the processing zone. Overall, this yields 
{\footnotesize
\begin{align*}
    \bigwedge_{0 < t \leq T} \bigwedge_{0 \leq j \leq |S|} \left(s^t_{j} \Rightarrow \bigwedge_{i_k \in C_j} x^t_{e_{in}, i_k} \right),
\end{align*}
}
where $|S|$ is the number of elements in the given sequence and $C_j$ is the set of ion chains that have to be moved to the processing zone because of element $j$ in the sequence. 

In order to formulate the order of the sequence, we process the sequence in a pairwise fashion. More precisely, if variable $s^t_j$ of sequence element $j$ is true at time step t, the variable of the next sequence element $j+1$ has to be true at a later time step $t'>t$.  Since every sequence element has a unique index, applying this to every variable enforces the correct order of sequence elements. The corresponding constraint is given by
{\footnotesize
\begin{align*}
    \bigwedge_{0 \leq j \leq L} \bigvee_{0 \leq t < T} \left(s^t_{j} \wedge  \bigvee_{t < t' \leq T}s^{t'}_{j+1} \right).
\end{align*}
}
Lastly, every sequence element $s^t_j$ is only true once. This leads to
{\footnotesize
\begin{align*}
    \bigwedge_{0 \leq j \leq L} \atmost \left( \left\{ s^t_j \right\}_{0 < t \leq T}\right).
\end{align*}
}
\begin{example}
    Consider again the sequence in \autoref{ex:sequence_ion_chains}: $[i_0; (i_0, i_1); (i_0, i_2); i_1; (i_1, i_2); i_3]$.
    Satisfying all discussed constraints, a valid solution given by the solver would start by moving ion chain $i_0$ to the processing zone. Both $i_1$ and $i_2$ then join $i_0$ one after the other. After they have been processed, both ion chains move back to the memory zone, while ion chain $i_1$ starts moving to the processing zone. To realize the rest of the sequence, $i_2$ joins $i_1$, before $i_3$ occupies the processing zone alone for at least one time step to finalize the sequence. 
    A valid assignment of all variables achieves these movements within a given amount of time steps $T$.
\end{example}

Overall, this results in a symbolic encoding of system states in a QCCD device and corresponding constraints to ensure validity in the position and movement of ion chains.
Consider a fixed number of time steps $T$.
If the SAT solver returns \enquote{unsat} (i.e.,~no satisfying assignment exists), then it has been proven that no movement realizing the given sequence within $T$ time steps exist. 
If the SAT solver determines a satisfying assignment, then such a movement within $T$ time steps has been obtained. 
Since $T$ is iteratively increased by 1 starting with $T=1$ (as described above in \autoref{sec:general-idea}), this eventually proves this movement to be minimal, i.e., $T=\hat{T}$. 
This symbolic encoding may also be extended to include new constraints that model the behavior of new types of QCCD devices.

\section{Empirical Evaluation}
\label{sec:experiments}

The approach proposed above has been implemented in Python3 utilizing the publicly available SAT solver Z3 (version 4.12.1)~\cite{10.5555/1792734.1792766}. The resulting implementation is available as open-source at \mbox{\url{https://github.com/cda-tum/mqt-ion-shuttler}}.
To the best of our knowledge, this leads to the first \emph{exact} shuttling approach for trapped-ion quantum computers. In this section, we discuss the efficacy of the resulting tool and summarize the obtained results.

To evaluate the tool, we considered different architectures following a grid structure, i.e.,~the angles of all junctions are set to \SI{90}{\degree}.
The corresponding grid-like graphs are denoted $L(m,n,v,h)$ and constructed as follows:
\begin{itemize}
	\item The graph is laid out as a $m \times n$ grid, with $m$ nodes vertically and $n$ nodes horizontally. 
	\item Each node in this grid represents a junction in the trapped-ion quantum computer.
	\item In between two major nodes (junctions) at most $v$ ion chains can be trapped vertically.
	\item In between two major nodes (junctions) at most $h$ ion chains can be trapped horizontally.
\end{itemize}
The grid is further extended by two additional edges which represent one outbound edge to a processing zone and one inbound edge leading back to the memory grid.
Using a random starting configuration of ion chains on these grids, we then used the proposed approach to determine the minimum number of time steps~$\hat{T}$ that are sufficient to realize a given quantum circuit, e.g., shuttling sequence.
All evaluations were conducted on a machine with an Intel(R) Xeon(R) W-1370P CPU (running at \SI{3.6}{\giga\hertz}) and \SI{128}{\gibi\byte} main memory running Python~3.8.10.

\begin{figure}[tbp]
\centering
    \begin{subfigure}{0.449\linewidth}
      \centering
      \includegraphics[trim=8mm 0 15mm 0, clip, width=.5\linewidth, angle=270]{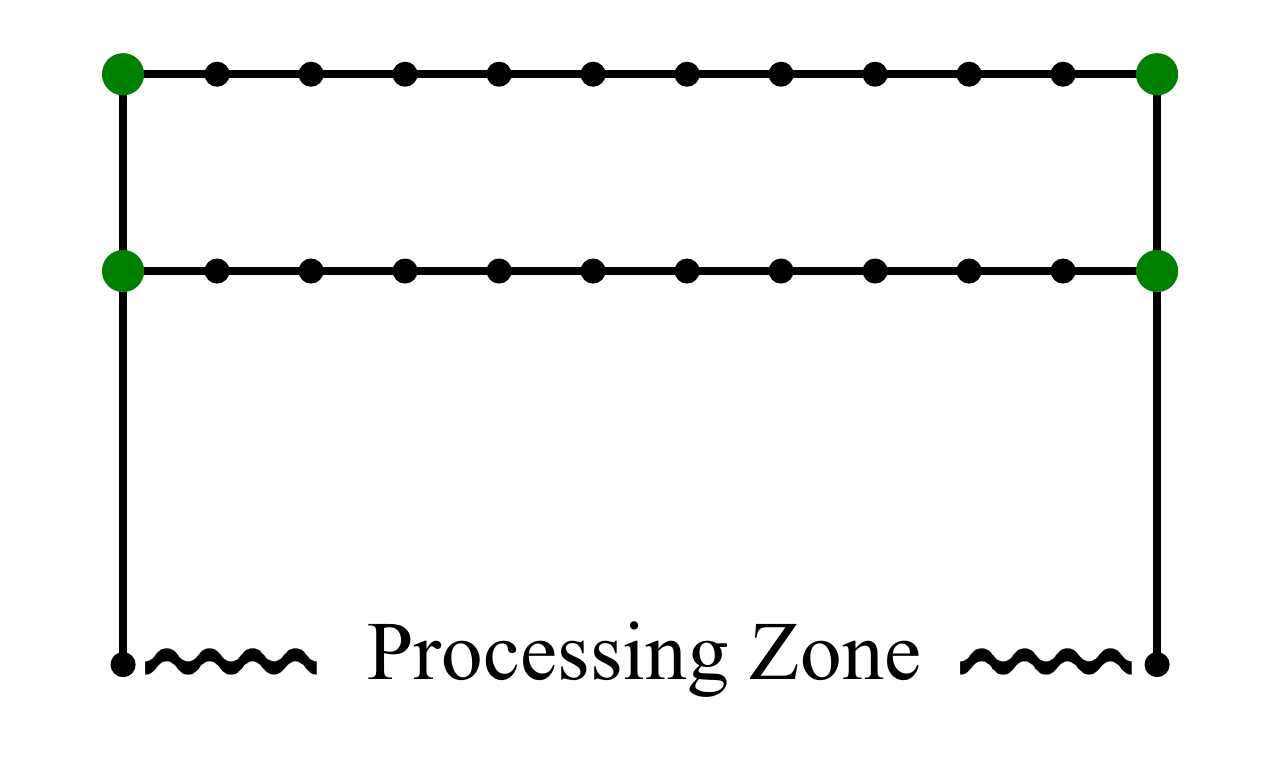}
      \caption{\enquote{Racetrack}: $L(2, 2, 1, 11)$}
      \label{fig:racetrack}
    \end{subfigure}\hspace{1em}%
    \begin{subfigure}{0.367\linewidth}
      \centering
      \includegraphics[trim=8mm 5mm 14.7mm 0, clip, width=.617\linewidth,angle=270]{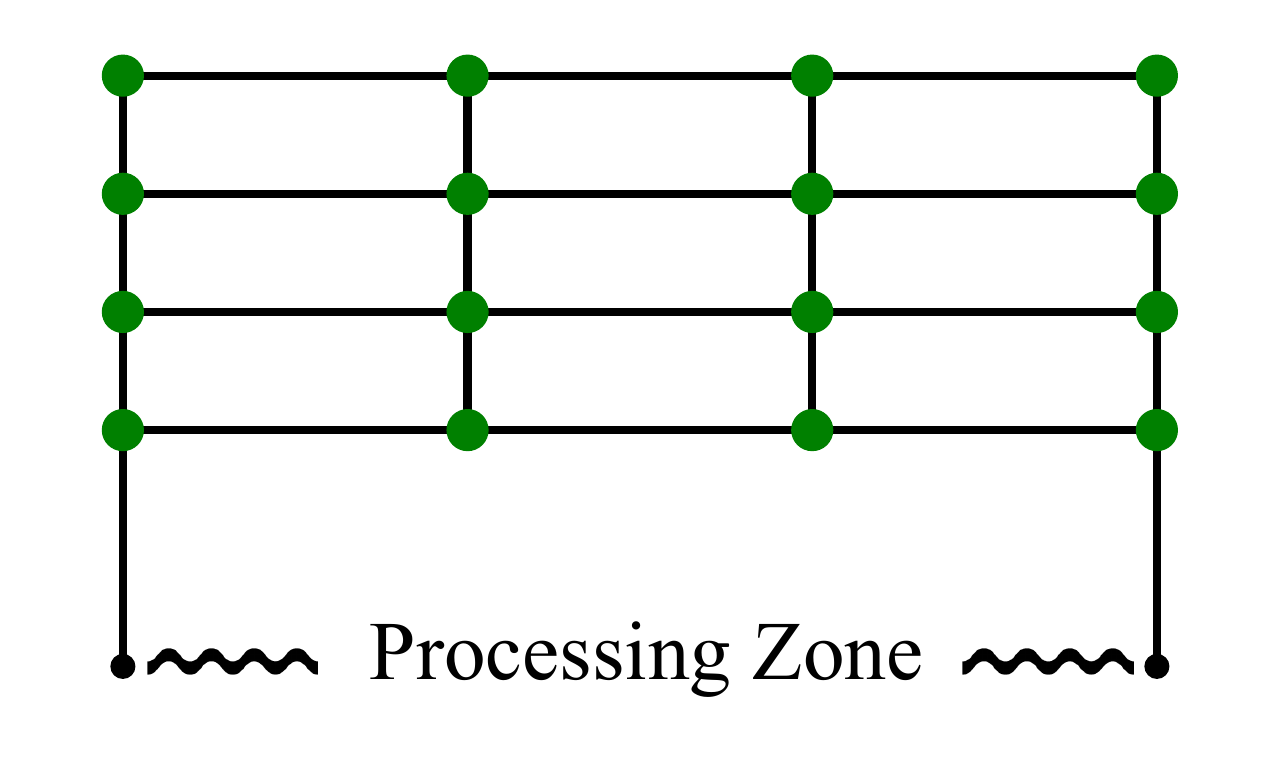}
      \caption{\enquote{Lattice}: $L(4, 4, 1, 1)$}
      \label{fig:grid}
    \end{subfigure}\hfill%
    \caption{Two types of layouts considered in the benchmark.}
    \label{fig:evaluation_layouts}
    \vspace{-1em}
\end{figure}

\begin{table*}[t]
    \centering
    \caption{Results of the Empirical Evaluation}
    \label{tab:benchmarks}
    \resizebox{\linewidth}{!}{\begin{tabular}{@{} c @{\hspace{2em}} c c r r r r @{\hspace{2em}} c c r r r r @{}}
        \toprule
        & \multicolumn{6}{c}{\textbf{Racetrack}} & \multicolumn{6}{c}{\textbf{Lattice}} \\
        \cmidrule(r{2em}){2-7} \cmidrule{8-13}
        Algorithm & $m$ $n$ $v$ $h$ & {$|C|$/$|E_M|$} & $|S|$ & {$\hat{T}$} & {$t_\textrm{CPU}$} & t.o. & $m$ $n$ $v$ $h$ & {$|C|$/$|E_M|$} & $|S|$ & {$\hat{T}$} & {$t_\textrm{CPU}$} & t.o. \\
        \midrule
        Full Register Access & 2 2 1 \phantom{1}5 & 6/12 (50\%) & 6 & 11.0 & 11.3 s & 0 & 3 3 1 1 & 6/12 (50\%) & 6 & 10.9 & 7.1 s & 0 \\
        & & 12/12 (100\%) & 12 & 22.6 & 1721.8 s & 0 
        & & 12/12 (100\%) & 12 & 17.0 & 36.1 s & 0 \\[3pt]
        & 2 2 1 11 & 6/24 (25\%) & 6 & 11.0 & 39.6 s & 0 & 4 4 1 1 & 6/24 (25\%) & 6  & 12.5 & 17.7 s & 0 \\
        & & 12/24 (50\%) & 12 & 21.4 & 1985.1 s & 1 & & 12/24 (50\%) & 12 & 18.5 & 80.6 s &0 \\
        & & 18/24 (75\%) & 18 & \textgreater28 & t.o. & 10 & & 18/24 (75\%) & 18 & 24.5 & 237.6 s & 0 \\[3pt]
        & 2 2 1 19 & 6/40 (15\%) & 6 & 10.9 & 106.0 s & 0 & 5 5 1 1 & 6/40 (15\%) & 6 & 12.9 & 31.1 s & 0 \\
        & & 12/40 (30\%) & 12 & 23.0 & 2404.7 s & 4 & & 12/40 (30\%) & 12 & 18.9 & 139.2 s & 0 \\
        & & 18/40 (45\%) & 18 & \textgreater28 & t.o. & 10 & & 18/40 (45\%) & 18 & 24.9 & 386.2 s & 0 \\[3pt]
        & 2 2 1 29 & 6/60 (10\%) & 6 & 10.0 & 97 s & 0 & 6 6 1 1 & 6/60 (10\%) & 6 & 14.3 & 58.3 s & 0 \\
        & & 12/60 (20\%) & 12 & 20.3 & 3616.0 s & 6 & & 12/60 (20\%) & 12 & 20.3 & 243.9 s & 0 \\
        & & 18/60 (30\%) & 18 & \textgreater25 & t.o. & 10 & & 18/60 (30\%) & 18 & 26.3 & 647.3 s & 0 \\
        \midrule
        Quantum Fourier Transform $q=5$ & 2 2 1 5 & 5/12 (42\%) &15 & 22.4 & 40.0 s & 0 & 3 3 1 1 & 5/12 (42\%) & 15 & 26.9 & 42.4 s & 0 \\
        \phantom{Quantum Fourier Transform }$q=6$ & & 6/12 (50\%) & 21 & 30.0 & 99.9 s & 0 & & 6/12 (50\%) & 21 & 33.9 & 90.6 s & 0 \\
        \phantom{Quantum Fourier Transform }$q=7$ & & 7/12 (58\%) & 28 & 38.6 & 221.2 s & 0 & & 7/12 (58\%) & 28 & 41.9 & 185.6 s & 0 \\
        \phantom{Quantum Fourier Transform }$q=8$ & & 8/12 (67\%) & 36 & 48.2 & 463.6 s & 0 & & 8/12 (67\%) & 36 & 50.9 & 378.7 s & 0 \\
        \bottomrule
    \end{tabular}}
\end{table*}

For this evaluation, we consider two categories of layouts. The first category contains layouts with long \enquote{streets} of connected edges that are not interrupted by junctions, similar to a \emph{racetrack}. We compared these layouts to \emph{lattice} layouts, in which every edge is enclosed by two junctions.
The major difference between these two types of layouts is the amount of junctions in the system. The racetrack configuration consists of the minimum amount of junctions for a grid-form graph, while the lattice type employs the maximum amount. One example of each category is given in \autoref{fig:evaluation_layouts}, where both layouts can trap up to 24 ion chains at a time.

Within the two different layouts, two different types of sequences were considered for the evaluation:
\begin{itemize}
    \item \emph{Demonstration of full register access:}
    In a first series, we investigated the accessibility of all registers in a system, i.e.,~the minimum number of time steps to move all ion chains in order to the processing zone. Hence, the corresponding sequence is an ascending list of all ion chain indices.
    This is akin to a quantum circuit, that applies a single-qubit gate successively to each qubit.
    \item \emph{Demonstration of quantum Fourier transform:}
    In a second series, we considered the sequence derived from the quantum Fourier transform~\cite{DBLP:books/daglib/0046438} (the precise instance has been taken from MQT Bench~\cite{Quetschlich_2023}).
    For better comparability, we set the amount of ions in each ion chain to one. Otherwise, the results would also strongly depend on this additional degree of freedom.
\end{itemize}
In all experiments, the initial placement of chains was done randomly.

The results are shown in \autoref{tab:benchmarks}. To this end, the results are split into one part concerning all racetrack layouts and the other all lattice layouts. 
In each block, the first column defines the variables of graph $L(m, n, v, h)$. Then, the second column provides the number of ion chains $|C|$, the number of edges that constitute the memory~zone~$|E_M|$ (for the sake of clarity, we only list the edges that vary in the considered layouts, i.e., all edges in the memory zone) and the percentage relationship between these two values.
The next three columns list the length of the sequence $|S|$, the minimal amount of time steps $\hat{T}$ obtained over $10$ runs, each with a different random starting configuration of ion chains, and the average time $t_{\textrm{CPU}}$ taken to generate these exact results.
The abbreviation \enquote{t.o.} indicates that no satisfying assignment has been determined for all $10$ runs within a timeout of~5\,000 CPU seconds (in these cases, the value on the $\hat{T}$-column provides the largest number of time steps for which a non-existence of a shuttling has been proven---providing a lower bound for $\hat{T}$).
If some, but not all runs resulted in a timeout, the averages of $\hat{T}$ and $t_{\textrm{CPU}}$ are given for the successful runs.
The last column provides the number of timeouts that occurred.
Variable~$q$ refers to the number of qubits in the considered quantum Fourier transform. 

Overall, the results confirm that the proposed approach is applicable to determine \emph{exact} results for shuttling in trapped-ion quantum computers. Although this is only possible for relatively small layouts (after all, the problem remains computationally expensive), this provides first-time \emph{minimal} solutions and, hence, lower bounds are available for this problem. 
The discussed approach enables the comparison of different layouts, providing a guideline for future proposals of QCCD architectures.
As an example, our first results indicate that, even for small instances, lattice layouts with many junctions provide better access to all registers compared to layouts with few junctions in respect to shuttling time. 

\section{Conclusion}
\label{sec:conclusion}

Quantum computers based on trapped-ion technology are promising candidates for enabling scalable quantum computations with the QCCD architecture.
Inside these architectures, efficient shuttling of ion chains in the memory zone is important to reduce run time and, thus, decrease the likelihood of errors.
In this paper, we proposed an abstraction of the physical memory zone in the QCCD architecture and a corresponding description by means of Boolean variables.
More precisely, we abstract the physical architecture through a graph, where nodes represent junctions and edges the individual sites where ion chains are held.
Based on this graph, we introduce Boolean variables to encode the position of the ion chains at any time step.
Added constraints ensure only valid states and transitions between states in a time step are possible.
Leveraging the power of modern SAT solver, we utilize the symbolic description to find exact solutions to the shuttling problem.
The empirical evaluation confirms the efficacy for determining exact shuttling.
Moreover, even in instances which timeout, i.e.,~do not finish in the given time, the results provide a lower bound for the number of required time steps.

\section*{\sc Acknowledgments}
This work was funded under the European Union's Horizon 2020 research and innovation programme (DA QC, grant agreement \mbox{No. 101001318} and MILLENION, grant agreement No. 101114305), the State of Upper Austria in the frame of the COMET program, the QuantumReady project within Quantum Austria (managed by the FFG), and was part of the Munich Quantum Valley, which is supported by the Bavarian state government with funds from the Hightech Agenda Bayern Plus.

\printbibliography
\end{document}